\documentclass[12pt]{article}
\usepackage{cite,amssymb,graphicx,graphics}
\begin{document}
\title{\bf Localization of Gauge Bosons in the 5D Standing Wave Braneworld}
\author{{\bf Merab Gogberashvili}\\
Andronikashvili Institute of Physics,\\
6 Tamarashvili St., Tbilisi 0177, Georgia \\
and\\
Javakhishvili State University,\\
3 Chavchavadze Ave., Tbilisi 0128, Georgia\\
{\sl E-mail: gogber@gmail.com} \\\\
{\bf Pavle Midodashvili}\\
Ilia State University, \\ 3/5 Kakutsa Cholokashvili Ave., Tbilisi 0162, Georgia\\
{\sl E-mail: pmidodashvili@yahoo.com} \\\\
{\bf Levan Midodashvili}\\
Gori University, \\ 53 Chavchavadze St., Gori 1400, Georgia \\
{\sl E-mail: levmid@hotmail.com}}
\maketitle
\begin{abstract}
We investigate the problem of localization of gauge fields within the 5D standing wave braneworld model and show that in the case of increasing warp factor there exist normalizable vector field zero modes on the brane.
\end{abstract}

\vskip 0.5cm


The brane models \cite{Hi,brane} have attracted a lot of interest recently with the aim of solving several open questions in modern physics. A key requirement for realizing the braneworld idea is that the various bulk fields be localized on the brane. For reasons of economy, stability of the model and avoidance of charge universality obstruction \cite{DuRuTi} one would like to have a universal gravitational trapping mechanism for all fields. However, there are difficulties to realize this mechanism with exponentially warped space-times. In the existing (1+4)-dimensional models spin $0$ and spin $2$ fields can be localized on the brane with the decreasing warp factor \cite{brane}, spin $1/2$ field can be localized with the increasing factor \cite{BaGa}, and spin $1$ fields are not localized at all \cite{Po}. In the case of 6D models it was found that spin $0$, spin $1$ and spin $2$ fields are localized on the brane with the decreasing warp factor and spin $1/2$ fields again are localized with the increasing factor \cite{Od}. There exist also 6D models with non-exponential warp factors providing gravitational localization of all kind of bulk fields on the brane \cite{6D}, however, these models require introduction of unnatural sources.

To solve the localization problem recently we had proposed the standing wave braneworld model \cite{Wave}, which is generated by collective oscillations of gravitational and scalar phantom-like fields (similar to \cite{GMS}) in 5D bulk. To avoid the well-known problems with stability one can associate the bulk ghost-like field with the geometrical scalar in a 5D integrable Weyl model \cite{Weyl}. In this model a massless scalar, either an ordinary or ghost-like scalar, appears through the generalization of the Riemannian definition of the covariant derivative of the metric tensor. The geometrical scalar fields have not dangerous couplings with matter fields, and it is known that the Weyl model is stable.

The metric of the model \cite{Wave} in the case of increasing warp factor has the form:
\begin{equation} \label{metric}
ds^2 = e^{2a|r|}\left( dt^2 - e^{u}dx^2 - e^{u}dy^2 - e^{-2u}dz^2 \right) - dr^2~.
\end{equation}
Here $a = \sqrt{\Lambda/6} > 0$ ($\Lambda$ is 5D cosmological constant) and
\begin{equation} \label{separation}
u(t,r) = B ~\sin (\omega t)~ e^{-2a|r|}~Y_2\left( \frac{\omega}{a} e^{-a|r|} \right)~,
\end{equation}
where $B$ is a constant, $\omega$ denotes the oscillation frequency of the standing wave, and $Y_2$ is the second-order Bessel function of the second kind.

The solution (\ref{metric}) describes the positive tension brane at $r = 0$, which undergoes anisotropic oscillations and sends waves into the 'sea' of phantom-like scalar field in the bulk. When the amplitude of anisotropy is small with respect to the tension of the brane (quasi-isotropic limit) effects of anisotropy will be observable only for cosmological distances. One can study brane cosmological model where anisotropy dissipates via inflation \cite{MSS}, or by leakage of thermal graviton radiation into the bulk \cite{NPK}.

Note that there is a large similarity between our solution and the simple electromagnetic standing wave between two infinite conducting planes. In our case there is only one plane -- the brane at $r = 0$. Instead of the second plane we have the wall which is provided by the static part of the gravitational potential increasing as one moves away from $r = 0$. This reminds the experiments of \cite{Nes} where neutrons were trapped between a reflecting plane at the surface of the Earth and the wall generated by the gravitational potential of the Earth. We have discrete frequencies of oscillation for our standing wave just as the neutrons had discrete energy levels in these experiments.

The $r$-dependent factor of the metric function (\ref{separation}) has finite number of zeros along the extra coordinate axis, which form the nodes of the standing wave. These nodes can be considered as 4D space-time 'islands', where the matter particles can be bound \cite{Wave}. One of nodes of the standing wave must be located at the position of the brane. This can be achieved by imposing the condition:
\begin{equation}\label{MetricFunctionAtOrigin}
Y_2\left( \frac{\omega}{a} \right)=0~,
\end{equation}
which fixes the ratio of the standing wave frequency $\omega$ to the curvature scale $a$.

If the frequency of oscillations in the bulk $\omega$ is much larger than frequencies associated with energies of matter particles on the brane standing waves can provide their localization. One of the classical analogs of this trapping mechanism is the so-called optical lattice with standing electromagnetic waves \cite{Opt}. The explicit localization of scalar and tensor field zero modes in standing wave braneworld with the increasing warp factor was demonstrated in our previous paper \cite{GMM}.

As mentioned above, the implementation of pure gravitational trapping mechanism of vector field particles on the brane in 5D space-time remains most problematic, although there was proposed some non-gravitational trapping mechanisms, for example \cite{DSh}. In this Letter we investigate the vector field localization problem and explicitly show that the standing wave braneworld metric (\ref{metric}) provides pure gravitational localization of vector field zero modes on the brane.


For simplicity in this Letter we investigate only the $U(1)$ vector field, the generalization to the case of non-Abelian gauge fields is straightforward. The action of vector field is:
\begin{equation}\label{VectorAction}
S = - \frac14\int d^5x\sqrt g~ g^{MN}g^{PR}F_{MP}F_{NR}~,
\end{equation}
where capital Latin indexes refer to 5D space-time,
\begin{equation} \label{F}
F_{MP} = \partial _M A_P - \partial _P A_M
\end{equation}
is the 5D vector field tensor, and $g$ is the determinant of the metric (\ref{metric}),
\begin{equation} \label{determinant}
\sqrt g = e^{4a|r|}~.
\end{equation}
The action (\ref{VectorAction}) gives the system of five equations:
\begin{equation}\label{VectorFieldEquations}
\frac{1}{\sqrt g }\partial _M\left( \sqrt g ~g^{MN}g^{PR}F_{NR} \right) = 0 ~.
\end{equation}

Let us seek for the solution to the system (\ref{VectorFieldEquations}) in the form:
\begin{eqnarray}\label{VectorDecomposition}
A_t(x^C) &=& \rho (r)~a_t(x^\nu)~, \nonumber \\
A_x(x^C) &=& e^{u(t,r)} \rho (r)~a_x(x^\nu)~, \nonumber \\
A_y(x^C) &=& e^{u(t,r)} \rho (r)~a_y(x^\nu)~, \\
A_z(x^C) &=& e^{-2u(t,r)} \rho (r)~a_z(x^\nu)~, \nonumber \\
A_r(x^C) &=& 0~,\nonumber
\end{eqnarray}
where $u(t, r)$ is the oscillatory metric function (\ref{separation}), $a_{\mu}(x^\nu)$ denote the components of 4D vector potential (Greek letters are used for 4D indices) and scalar factor $\rho(r)$ depends only on the extra coordinate $r$. The last expression in (\ref{VectorDecomposition}) in fact is the 5D gauge condition.

We require existence of flat 4D vector waves localized on the brane,
\begin{equation} \label{Factors}
a_\mu\left(x^\nu\right) \sim \varepsilon_\mu e^{i(Et + p_x x + p_y y + p_z z)} ~,
\end{equation}
where $E$, $p_x$, $p_y$, $p_z$ are components of energy-momentum along the brane. Solutions of this kind exist only on the brane, where $u = 0$, and in the case when $\omega$ is much larger than frequencies associated with the energies of the particles on the brane,
\begin{equation}
\omega \gg E~.
\end{equation}
For such high frequencies of bulk standing waves we can perform time averaging of oscillatory functions in (\ref{VectorFieldEquations}).

Taking into account the equalities for time averages \cite{GMM}:
\begin{equation}\label{AdditionalFacts}
\left\langle u \right\rangle  = \left\langle
u' \right\rangle= \left\langle \frac{\partial
u}{\partial t} \right\rangle  =\left\langle \frac{\partial
u}{\partial t}e^{ - u} \right\rangle  = 0~,
\end{equation}
where the prime denotes the derivative with respect to the extra coordinate $r$, time averaging of the fifth ($r$-component) equation of the system (\ref{VectorFieldEquations}),
\begin{equation}
\partial_\alpha \left( g^{\alpha\beta}  A_\beta '\right) = 0~.
\end{equation}
gives the Lorenz-like gauge condition:
\begin{equation} \label{GaugeCondition2}
g^{\alpha\beta}\partial_\alpha A_\beta =
\eta^{\alpha\beta}\partial_\alpha a_\beta = 0~,
\end{equation}
where $\eta_{\alpha\beta}$ denotes the metric of 4D Minkowski space-time. The equation (\ref{GaugeCondition2}) together with the last expression of (\ref{VectorDecomposition}) can be considered as the full set of imposed gauge conditions.

Remaining four equations of the system (\ref{VectorFieldEquations}),
\begin{equation}
\partial _\gamma\left( g^{\gamma\delta}g^{\beta\alpha}F_{\delta\alpha} \right) - \frac{1}{\sqrt g }\left( \sqrt g ~g^{\beta\alpha}A'_\alpha \right)' = 0 ~,
\end{equation}
after time averaging and using of (\ref{AdditionalFacts}) and (\ref{GaugeCondition2}) reduce to:
\begin{equation} \label{system-a}
\rho ~g^{\alpha \delta}\partial _\alpha\partial _\delta a_\beta + e^{-2a|r|} \left( e^{2a|r|}\rho' \right)' a_\beta = 0 ~.
\end{equation}
In the case of the zero mode (\ref{Factors}),
\begin{equation}
E^2 - p_x^2 - p_y^2 - p_z^2 = 0~,
\end{equation}
the system (\ref{system-a}) gives the single equation for $\rho(r)$:
\begin{equation}\label{GeneralEquation}
\left( e^{2a|r|}\rho' \right)' - K(r) \rho  = 0~,
\end{equation}
where
\begin{equation} \label{K(r)}
K(r) = \left(\left\langle e^{-u} \right\rangle -1\right)\left( p_x^2 + p_y^2 \right) + \left(\left\langle e^{2u} \right\rangle -1\right)p_z^2 ~.
\end{equation}
Time averages of exponential metric functions were calculated in \cite{GMM} and have the form:
\begin{equation}\label{TimeAverafes}
\left\langle e^{bu} \right\rangle = I_0 \left( \frac{|b B| a^2 z}{\omega ^2} \left[ 2 Y_1 \left(z\right)-z Y_0\left(z\right) \right] \right),
\end{equation}
where $b$ is a constant, $I_0$ is the modified Bessel function of the zero-order, and
\begin{equation}
z =\frac{\omega}{a}~ e^{-a|r|}~.
\end{equation}

By making the change:
\begin{equation}\label{FunctionChange}
\rho(r) = e^{-a|r|}\psi(r)~,
\end{equation}
we put the equation (\ref{GeneralEquation}) into the form of a non-relativistic quantum mechanical problem:
\begin{equation}\label{AnalogEquation}
\psi''(r)-U(r)\psi(r) = 0~,
\end{equation}
where the potential:
\begin{equation}\label{AnalogPotential}
U(r)= 2a\delta(r) + a^{2} + e^{-2a|r|}K(r) ~,
\end{equation}
is continuous function. FIG. 1 shows behavior of $U(r)$ for the first zero of the Bessel function $Y_{2}$, i.e. when
\begin{equation}\label{FirstZerosY}
\frac{\omega}{a} \approx 3.38~.
\end{equation}

\begin{figure}
\begin{center}
\includegraphics[width=0.35\textwidth]{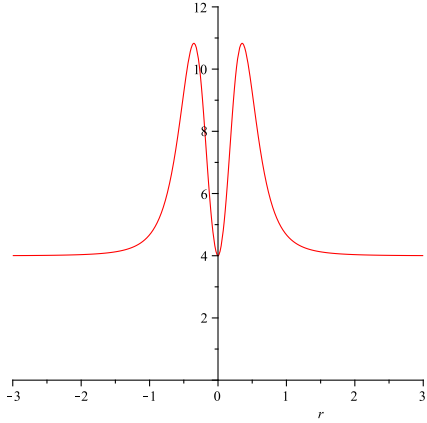}
\caption {The total gravitational potential (\ref{AnalogPotential}) for vector particles in the bulk.}
\end{center}
\end{figure}

To determine general behavior of extra dimension factor of the vector field zero modes, $\rho(r)$, we explore Eq. (\ref{AnalogEquation}) in the two limiting regions: far from and close to the brane. The function $K(r)$, which takes into account oscillatory properties of the standing wave, has the following asymptotical forms:
\begin{eqnarray}
\left. K (r)\right|_{r \to 0} &\sim& r^2~, \nonumber \\
\left. K (r)\right|_{r \to \infty} &\sim& const~.
\end{eqnarray}
So close to the brane Eq. (\ref{AnalogEquation}) obtains the form:
\begin{equation}\label{phi}
\psi '' - \left[ 2a \delta (r) + a^2 \right]\psi = 0~,
\end{equation}
with the solution:
\begin{equation} \label{phi-0}
\left. \psi (r)\right|_{r \to 0} \sim Ce^{a|r|}~,
\end{equation}
where $C$ is a constant. At the infinity Eq. (\ref{AnalogEquation}) again reduces to (\ref{phi}) (but without delta-function of course) with the two solutions,
\begin{equation} \label{phi-infinity}
\left. \psi (r)\right|_{r \to \infty} \sim e^{\pm a|r|}~.
\end{equation}

Resemblance of asymptotic forms of Eq. (\ref{AnalogEquation}) is not surprising because of the form of the potential (\ref{AnalogPotential}) (see FIG. 1). Standing waves in the bulk are bounded by the brane at $r=0$ and by the gravitational potential at the infinity, i.e. these two regions correspond to the nodes of the waves.

To have normalizable vector zero mode we impose the boundary conditions:
\begin{eqnarray}
\left. \rho' (r)\right|_{r=0} &=& 0~, \nonumber \\
\left. \rho (r)\right|_{r\to \infty} &=& 0~.
\end{eqnarray}
Then using the definition (\ref{FunctionChange}) and the solutions (\ref{phi-0}) and (\ref{phi-infinity}) we find:
\begin{eqnarray}
\left. \rho (r)\right|_{r \to 0} &\sim& C ~, \nonumber \\
\left. \rho (r)\right|_{r\to \infty} &\sim& e^{-2a|r|}~.
\end{eqnarray}
So $\rho (r)$ has maximum on the brane and falls off at the infinity as $e^{-2a|r|}$. For such extra dimension factor the integral over $r$ in the action (\ref{VectorAction}) is convergent, what means that 4D vector fields are localized on the brane.

Indeed, using the {\it ansatz} (\ref{VectorDecomposition}) and the equalities (\ref{AdditionalFacts}), time-averaged components of the vector field tensor (\ref{F}) can be written in the form:
\begin{eqnarray} \label{FF}
\left\langle F_r^\nu \left(x^C \right)\right\rangle &=& \left\langle g^{\nu\beta}F_{r\beta} \right\rangle = e^{-2a|r|}\rho' ~a^\nu \left(x^\alpha \right)~, \nonumber \\
\left\langle F_\nu^\mu \left(x^C \right) \right\rangle &=& \left\langle g^{\nu\beta}F_{\mu\beta} \right\rangle = e^{-2a|r|}\rho \left[ f_\nu^\mu \left(x^\alpha \right) + M_\nu^\mu \left(\left\langle e^{bu} - 1 \right\rangle \partial_\beta a^\alpha \right)\right]~,
\end{eqnarray}
where
\begin{equation}
f_{\mu\nu} \left(x^\alpha \right) = \partial _\mu a_\nu - \partial _\nu a_\mu
\end{equation}
is the 4D vector field tensor and $M_\nu^\mu $ represents the sum of terms of the type $\left\langle e^{bu} - 1 \right\rangle \partial_\nu a^\mu$. Since the functions $\rho'$ and $\left\langle e^{bu} - 1 \right\rangle $ in (\ref{FF}) vanish on the brane, the 5D vector Lagrangian at $r = 0$ gets the standard 4D form:
\begin{equation}\label{Lagrangian}
L = - \left. \frac14\sqrt g~ \left\langle F_M^P \right\rangle \left\langle F_P^M \right\rangle \right|_{r=0} = -\frac14 C^2 f_\alpha^\beta  f_\beta^\alpha ~.
\end{equation}
Of course in general the 5D Lagrangian of vector fields is more complicate. For instance, it contains the mass term $\left(\rho' F_r^\alpha\right)^2 \sim \left(\rho'\right)^2 a_\alpha a^\alpha$, i.e. the zero mode vector particles are massless only on the brane and acquire masses in the bulk. This fact can be considered as the alternative mechanism of localization.


To conclude, in this Letter we have explicitly shown the existence of normalizable zero modes of gauge bosons on the brane in the 5D standing wave braneworld model with increasing warp factor.

\medskip


\noindent {\bf Acknowledgments:} The research was supported by the grant of Shota Rustaveli National Science Foundation $\#{\rm GNSF/ST}09\_798\_4-100$.


\end{document}